\documentclass[10pt,letterpaper]{article}
\usepackage{t1enc}
\usepackage[latin1]{inputenc}
\usepackage[english]{babel}
\usepackage{graphics}
\begin{document}
\date{}
\title{ On the Jacobi Elliptic functions and Applications}
\author{A.  Raouf  Chouikha \footnote
{Universite Paris 13 LAGA UMR 7539 Villetaneuse 93430,  e-mail: chouikha@math.univ-paris13.fr}
}
\maketitle

\begin{abstract}
In this paper we are interested in developments of elliptic functions of Jacobi.
In particular a trigonometric expansion of the classical theta functions introduced by the author (Algebraic methods and q-special functions, Editors: C.R.M. Proceedings  and Lectures Notes, A.M.S., vol 22,   Providence, 1999, 53-57) permits one establish a differential system. This system is derived from the heat equation and is satisfied by their coefficients. Several applications may be deduced.\\ Other types of expansions for the Jacobi elliptic functions as well as for the Zeta function are examined. 
\end{abstract}

\section{Introduction}

We review briefly some known facts on Jacobi elliptic functions theta and zeta functions for later use.
(For details see, e.g. [1], [2].)\\
Let $\theta $ be the temperature at time $t$ at any point in a solid the conducting properties of which are uniform and isotropic. If $\rho $ is its density, $s$ its specific heat and $k$ its thermal conductivity, $\theta $ satisfies the heat equation :
$$\kappa \nabla^2\theta = \frac {\partial \theta }{\partial t},$$
where $\kappa = \frac {k}{s\rho}$ is the diffusivity. \\ Let $Ouvw$ be a rectangular Cartesian frame. In the special case where there is no variation of temperature in the $uw-$plane, the heat flow is everywhere parallel to the $v-$axis and the heat equation reduces to the form 
\begin{equation}
\kappa\frac {\partial ^2 y}{\partial v^2} = \frac {\partial y}{\partial t }
\end{equation}
where $y = \theta (v,t)$.\\

Consider the following boundary conditions
$$\theta (0,t) = \theta (1,t), \quad \theta (v,0) = \pi \delta (v-1/2), \quad 0 < v < 1,$$
where $\delta (v) $ is its Dirac  function. Then the solution of the boundary value problem is given by 
\begin{equation}
\theta (v,t) = 2 \sum_{n\geq 0} (-1)^n e^{-(2n+1)^2 \pi^2 \kappa t} \sin ((2n+1) \pi v),
\end{equation} 
Consider the change  $$\tau = 4 i \pi \ \kappa t.$$
It follows that $\frac {1}{\kappa} \frac {\partial y}{\partial t} = 4 i \pi \frac {\partial y}{\partial \tau} $ and Equation (1.1) becomes the partial differential equation 
\begin{equation}
\frac {\partial ^2 y}{\partial v^2} = 4 i \pi \frac {\partial y}{\partial \tau }
\end{equation}
When we write \ $q = e^{i \pi \tau} = e^{-4 \pi^2 \kappa t},$ the solution (1.2) takes the form
\begin{equation}
\theta _1(v,\tau ) = 2 \sum_{n\geq 0} (-1)^n q^{(\frac{n+1}{2})^2} \sin ((2n+1) \pi v)
\end{equation}
which is the first of the four theta functions of Jacobi.\\
When the precise value of $q$ is not important, we suppress the dependance upon $q$.
If one changes the boundary conditions to
$$\frac {\partial \theta}{\partial v } = 0\ on \ v = 0, v = 1, \theta (v,0) = \pi \delta (v-1/2), \quad 0 < v < 1,$$
then the corresponding solution of the boundary value problem of the heat equation (1.1) is given by 
\begin{equation}
\theta_4(v) = \theta _4(v,\tau ) = 1 + 2 \sum_{n\geq 1 } (-1)^n q^{n^2} \cos (2n \pi v).
\end{equation}

The function \ $\theta _1(v,\tau )$ \  is periodic with period $2$. Incrementing  $v$ by $1/2$ it yields the second theta function  
\begin{equation}
\theta_2(v) = \theta _2(v,\tau ) = 2 \sum_{n\geq 0} q^{(\frac{n+1}{2})^2} \cos ((2n+1) \pi v).
\end{equation}
Similarly, the increment of $v$ by $1/2$ for $\theta _4(v,\tau )$  yields the third theta function
\begin{equation}
\theta_3(v) = \theta _3(v,\tau ) = 1 + 2 \sum_{n\geq 1} q^{n^2} \cos (2n \pi v).
 \end{equation}
 
 It is known the four theta functions $\theta_1, \theta_2, \theta_3, \theta_4$ can be extended to complex values for $v$ and $q$ such that $\mid q\mid  < 1$.\\
Note that Jacobi's fundamental work on the theory of elliptic functions was based on these \newpage four theta functions. His paper "Fundamenta nova theoria functionum ellipticarum'' published in 1829, together with its later supplements, made fundamental contributions to the theory of elliptic functions. \\

 Turn now to the Jacobi elliptic functions \ $sn\ u,\ cn\ u,\ dn\ u.$ \ There are defined as ratios of theta functions 
\begin{equation}
sn\ u = \frac {\theta_3(0)\theta_1(v)  }{\theta_2(0) \theta_4(v)}, \quad cn\ u = \frac {\theta_4(0)\theta_2(v)  }{\theta_2(0) \theta_4(v)},\quad dn\ u = \frac {\theta_4(0)\theta_3(v)  }{\theta_3(0) \theta_4(v)},
\end{equation}
where \ $u = \theta_3^2(0) v$.\\
Define parameters $k$ and $k'$ by 
$$k = \frac {\theta_2^2(0)}{\theta_3^2(0)}, \qquad k' = \frac {\theta_4^2(0)}{\theta_3^2(0)}.$$
They are called the modulus and the complementary modulus of the elliptic functions. When it is required to state the modulus explicitly, the elliptic functions of Jacobi are written \ $sn\ (u,k), \ cn\ (u,k),\ dn\ (u,k).$\\
Moreover as for the theta functions the three Jacobi elliptic functions are related. In particular they satisfy the following relations
\begin{equation}
sn^2\ u + cn^2\ u = 1 , \quad dn^2\ u + k^2 sn^2\ u = 1 , \quad k^2 cn^2\ u + k'^2 =  dn^2\ u .
\end{equation} 
\begin{equation}
sn'\ u = (cn\ u) (dn\ u), \ cn'\ u = -(sn\ u) (dn\ u), \ dn'\ u = -k^2 (sn\ u) (cn\ u).
\end{equation}\ 
The functions \ $sn\ (u,k),\ cn\ (u,k),\ dn\ (u,k)$ \ are doubly periodic with periods  $$(4K(k),i2K'(k)), (4K(k),2K(k)+i2K'(k), (2K(k),i4K'(k)$$ respectively. Here \ $K(k)$\ denotes the complete elliptic integral of the first kind $$K = 2\int_0^{\frac {\pi }{2}}\frac {dx}{\sqrt {1 - k^2 \sin ^2x}}$$ and \ $K'(k) = K(1-k)$.\ 
The modulus is such that \ $0 < k < 1$.\\
The limit case $k = 0$ yields $K(0) = \frac {\pi}{2}$ and trigonometric functions:  $$sn\ (u,0) = \sin u, \ cn\ (u,0) = \cos u, \ dn\ (u,0) = 1.$$
The limit case $k = 1$ yields $K(1) = \infty$ and hyperbolic functions: $$sn\ (u,1) = \tanh u, \ cn\ (u,1) = sech\ u, \ dn\ (u,1) = sech\ u.$$

The Zeta function of Jacobi is defined by 
$$Z(u) = \frac {d}{du}[Ln (\theta_4(v))], \qquad  u = \theta_3^2(0) v$$
and satisfies the following identity
$$Z(u+w) = Z(u) + Z(w) - k^2 (sn\ u)(sn\ w)(sn\ (u+w)).$$
\newpage
One demonstrated in [1] a new type of trigonometric development for theta functions. This one is of course connected to the developments of classic type. \\ 
Thanks to the heat equation we deduced  modular and arithmetic properties of its coefficients that seem to be of interest. \\  
Firstly we briefly recall significant results of [1] and [3]. The proofs are omitted. In light of these results one examine thereafter properties of elliptic and Zeta functions of Jacobi . 

\section{Theta functions}
We proved the next result

\bigskip
	
 {\bf Theorem 1}\qquad {\it The theta function\ $\theta_4(v,\tau)$\  may be expressed under the form $$\theta_4(v,\tau) = \theta_4(0,\tau)\ \exp\big[ \sum_{p\geq 1}
	c_{2p}(\tau) (\sin\pi v)^{2p}\big]$$
	  where the coefficients \ $c_{2p} $ \ verify the recurrence relation for \ $p \geq 1$ 
	$$ (A)
 \quad \cases{
4 ! \ {2p+4 \choose 4}\ c_{2p+4}  =  (2p+1)(2p+2) \big [(2p+2)(2p+3) + 4 p^2 - c_0 \big ]c_{2p+2} & \cr
  + (2p)^2 [ c_0 - (2p)^2 ] c_{2p} - 6 \big [ (2p+1)(2p+2) c_{2p+2} - 2c_2 - \sum_{k=1}^p 2k c_{2k} \big ]^2  & \cr}$$
	and \quad  $ c_0 = - 4 [\theta_2^4(0,\tau) + \theta_3^4(0,\tau)], \quad  c_2 = \frac {1}{2 \pi ^2}  \frac {\theta ''_4 (0,\tau )}{\theta _4 (0,\tau ) } $  and  $ c_4 =  \frac {1}{3} \theta_2^4(0,\tau)  \theta_3^4(0,\tau) + \frac {1}{3} c_2.$
	
	Moreover, the expression above for  \ $\theta_4$\ is valid in the strip}  $\mid Im v\mid < \frac {1}{2} Im \tau $.

	\bigskip
	
	For the other theta functions we obtain the following

	\bigskip
	
 {\bf Theorem 2}\qquad {\it 
 Under the hypotheses of  Theorem  1 we get the following expressions 
$$\theta_1(v,\tau) = \theta_4(0,\tau)\ exp[i\pi( v + {1\over 4}\tau) + 
\sum_{p\geq 1} c_{2p}(\tau) \sin^{2p}\pi (v+{1\over 2}\tau)]$$  
$$\theta_2(v,\tau) = \theta_4(0,\tau)\ exp[i\pi( v + {1\over 4}\tau) + 
\sum_{p\geq 1} c_{2p}(\tau) \cos^{2p}\pi (v+{1\over 2}\tau)]$$ 
$$\theta_3(v,\tau) = \theta_4(0,\tau)\ \exp\big[ \sum_{p\geq 1}
 c_{2p}(\tau) (\cos\pi v)^{2p}\big],$$
where coefficients \ $c_{2p}$ \ verify relation (A).\\
Moreover, the expressions above of \ $\theta_1$\ and \ $\theta_3$\ are valid in the strip \ $\mid Im v\mid <   Im \tau  $\ and, \ $\theta_2$\ is valid in the strip} $\mid Im v\mid < \frac {1}{2}  Im \tau  $.

 \newpage

 Under the same hypotheses, 
 the product of theta functions holds
$$\frac {\theta_2(v,\tau) \theta_3(v,\tau) \theta_4(v,\tau)}{\theta_4^3(0,\tau)} =  e^{v+{\tau \over 4}} \ \exp \big[\sum_{p\geq 1} c_{2p}(\tau) \big[ \sin^{2p}\pi v + \cos^{2p}\pi v + \cos^{2p}\pi (v+{1\over 2}\tau)\big] \big].$$
In particular we get 
$$\theta_1'(0,\tau) = \pi \theta_4^3(0,\tau) q^{1\over 4}\  \exp \big[\sum_{p\geq 1} c_{2p}(\tau) \big[ 1 + \cos^{2p}\pi {\tau \over 2}\big] \big].$$

The heat equation permits one to state a differential system satisfied by coefficients \  
$c_{2p}(\tau )$ \  
\begin{equation}
(S) \quad \cases{
\frac {4}{\pi} c'_{2p} = (2p+2)(2p+1)\ c_{2p+2} - 4 p^2\ c_{2p} -  & \cr 
 4 {\displaystyle \sum_{m=0}^{p-1}}  m \ c_{2m}[  (p-m) \ c_{2p-2m} -  (p-m+1) \ c_{2p-2m+2}]. & \cr}
\end{equation}
where \ $c'_{2p} = \displaystyle{\frac {d  c_{2p}}{d \tau }}$.
More precisely this system is obtained by identification after replacing 
expression of  \ $\theta_4(v,\tau) = \theta_4(0,\tau)\ \exp[ \sum_{p\geq 1}
	c_{2p}(\tau) (\sin\pi v)^{2p}]$ \  in Equation (1.3).\\
	
 The next theorem solves System (S) and thus an expansion of theta function is derived

\bigskip

	{\bf Theorem 3}\qquad {\it The coefficients  \  
$c_{2p}(\tau )$ \   may be expressed as
$$c_{2p}(\tau ) = - \frac {1}{p} \sum_{k\geq 0} \frac {1}{(\sin (k+\frac {1}{2})\pi \tau)^{2p}} = - \frac {1}{p} \sum_{k\geq 0} \bigg [\frac {(-4) q^{2k+1} }{(1 - q^{2k+1})^2}\bigg ]^p.$$
The function \ $\theta_4 $\ has the following expansion
$$\theta_4(v,\tau) = \theta_4(0,\tau)\ \exp\big[- \sum_{p\geq 1} \sum_{ k\geq 0} \frac {1}{p} \bigg( \frac {\sin \pi v}{(\sin (k+\frac {1}{2})\pi \tau)}\bigg)^{2p}\big].$$
Moreover the expression above of \ $\theta_4 $ \ is valid in the strip} \ $\mid Im v \mid < \frac {1}{2}  Im \tau  .$

	\bigskip
	
Of course the other theta functions \ $\theta_1(v,\tau), \theta_2(v,\tau), \theta_3(v,\tau) $\ have similar trigonometric expansions.
 
\section{Elliptic functions of Jacobi}
In this section we  introduce new trigonometric developments for Jacobi elliptic functions constructed from the theta functions. 
\newpage
	 {\bf Theorem 4}\qquad {\it Let \  $u = \theta_3^2(0) v$, \ such that $\mid Im v\mid < \frac {1}{2} Im \tau $.\ Then  the following expansions for elliptic functions hold }
	 $$sn\ u = e^{i\pi v}\ \exp [-\sum_{p\geq 1} \frac {1}{p} \sum_{k\geq 0} \frac {1}{(\sin (k+\frac {1}{2})\pi \tau)^{2p}} [ \sin^{2p}\pi (v+\frac {\tau}{2}) + \sin^{2p}\pi v - \cos^{2p}\pi \frac {\tau}{2} + 1]\ ] ,$$
	 $$cn\ u = e^{-i\pi v}\ \exp [-\sum_{p\geq 1} \frac {1}{p} \sum_{k\geq 0} \frac {1}{(\sin (k+\frac {1}{2})\pi \tau)^{2p}} [\sin^{2p}\pi v - \cos^{2p}\pi (v+\frac {\tau}{2}) - \cos^{2p}\pi \frac {\tau}{2}]\ ] ,$$
$$dn\ u = \exp [-\sum_{p\geq 1} \frac {1}{p} \sum_{k\geq 0} \frac {1}{(\sin (k+\frac {1}{2})\pi \tau)^{2p}} [ \cos^{2p}\pi v - \sin^{2p}\pi v - 1]\ ] .$$

	\bigskip
	
{\bf Proof}\qquad  
By Theorem 2 we get also the following expressions for ratios of theta functions
$$\frac {\theta_1(v,\tau)}{\theta_2(v,\tau)} = \exp [\sum_{p\geq 1} c_{2p}(\tau) [ \sin^{2p}\pi (v+{\tau\over 2}) - \cos^{2p}\pi (v+{\tau\over 2})] ]$$
$$\frac {\theta_3(v,\tau)}{\theta_4(v,\tau)} = \exp [\sum_{p\geq 1} c_{2p}(\tau) [ \cos^{2p}\pi v - \sin^{2p}\pi v ] ] .$$
The result follows by Theorem 3 since \ $c_{2p}(\tau) = -\frac {1}{p} \sum_{k\geq 0} (\sin (k+\frac {1}{2})\pi \tau)^{-2p}.$\\  Moreover by definition one has  \ ${\displaystyle  \frac {sn\ u}{cn\ u} = \frac {\theta_3(0)\theta_1(v,\tau)}{\theta_4(0)\theta_2(v,\tau)}}$\ and \\ ${\displaystyle 
sn\ u = \frac {\theta_3(0)\theta_1(v)  }{\theta_2(0) \theta_4(v)}, \quad cn\ u = \frac {\theta_4(0)\theta_2(v)  }{\theta_2(0) \theta_4(v)},\quad dn\ u = \frac {\theta_4(0)\theta_3(v)  }{\theta_3(0) \theta_4(v)}.}$

	\bigskip

Starting from Theorem 4 other various relations may be deduced.

	\bigskip
	
 {\bf Theorem 5} \qquad {\it  Under hypotheses of Theorem 4 the following relations hold}
$$\frac {sn\ u}{cn\ u} = \exp \big[-\sum_{p\geq 1} \frac {1}{p} \sum_{k\geq 0} \frac {1}{(\sin (k+\frac {1}{2})\pi \tau)^{2p}} \big[1 + \sin^{2p}\pi (v+{1\over 2}\tau) - \cos^{2p}\pi (v+{1\over 2}\tau)\big] \big].$$
$$\frac {\partial sn\ u}{\partial u} = e^{-i\pi v}\ \exp \big[-\sum_{p\geq 1} \frac {1}{p} \sum_{k\geq 0} \frac {1}{(\sin (k+\frac {1}{2})\pi \tau)^{2p}} \big[\cos^{2p}\pi v - \cos^{2p}\pi (v+\frac {\tau}{2}) - \cos^{2p}\pi \frac {\tau}{2} - 1\big]\ \big] .$$

	\bigskip

By the same way one obtains expansions for partial derivatives of \ $cn\ u$\ and \ $dn\ u$.\\

\section{Zeta function}
Consider the zeta function of Jacobi. It is defined by 
$$Zn(z,k) =  \frac {1}{2K} \frac {d}{dz} \log \theta _4(v,\tau ),$$\newpage   where \ $v = \frac {z}{2K}$\ and \ $K = 2\int_0^{\frac {\pi }{2}}\frac {dx}{\sqrt {1 - k^2 \sin ^2x}}$\ is the complete elliptic integral of the first kind and the modulus is such that \ $0 < k < 1$.
Note that the zeta function has also a Fourier expansion 

$$Zn(z,k) = \frac {2\pi}{K} \sum_{n\geq 1}\frac {q^n}{1-q^{2n}} \sin \frac {n \pi z}{K} .$$
which may be rewritten 
$$Zn(z,k) = \frac {\pi}{2K} \sin (\pi 2v) \sum_{k\geq 0} \frac {1}{\sin ^2(\pi v) - \sin ^2(k+\frac {1}{2}\pi \tau)}$$
where \ $v = \frac {z}{2K}.$
 
 	\bigskip 
 	
 {\bf Theorem 6} \qquad {\it  Let\  $K = 2\int_0^{\frac {\pi }{2}}\frac {dx}{\sqrt {1 - k^2 \sin ^2x}}$\ be the complete elliptic integral of the first kind. \\
 The zeta function of Jacobi has the following form
$$Zn(z,k) =  \frac {\pi}{2K} \sin (\pi \frac {z}{K}) \sum_{k\geq 0} \sum_{p\geq 1}  \bigg( \frac {\sin \pi \frac {z}{2K}}{(\sin (k+\frac {1}{2})\pi \tau)}\bigg)^{2p}$$
which is valid in the strip}\  $\mid Im \frac {z}{2K}\mid < \frac {1}{2} Im \tau $.

	\bigskip

\section{Concluding remarks}
The Jacobi elliptic functions and in particular \ $dn\ (u,k)$\ play an important role in the theory of elliptic functions as well as in many physical problems.\\
The previous calculations particularly indicate to us that the theory of the Jacobi elliptic functions seems not to be exhausted completely and new characterizations involving Jacobi theta functions may be found.  So we may expect always to discover other properties having interesting applications, 
as the works of Khare, Lakshminarayan and Sukhatme [5].\\  

We recall some quantum mechanical facts using the function \ $dn\ (u,k)$.\\ The wave functions \ $\psi_0^{\pm} = [dn\ (u,k)]^{\mp}$\ are the zeros modes of the periodic supersymmetric partners potentials : 
$$V_+(u) = \frac {2-k+2(k-1)}{dn^2\ (u,k)}\quad {\mbox and} \quad V_-(u) = 2 - k + 2 {dn^2\ (u,k)}.$$
This function also allows a resolution of a nonlinear Schrodinger equation. Indeed the nonlinear Schrodinger equation 
$$\frac {\partial \psi}{\partial t} + \frac {\partial^2 \psi}{\partial x^2} + 2 \psi^2 \bar{\psi} = 0$$
has the following as general periodic solutions
$$\psi(x,t) = r\ \exp[i(px-p^2-(2-k^2)r^2)t]\ dn\ (rx-2prt,k^2),$$
where $r$ and $p$ are some constants and $k$ is the elliptic modulus. The cyclic identities as well as their generalized Landen formulas play an important role in showing that a kind of linear superposition of periodic solutions is valid in physically interesting nonlinear differential equations. 
See [5] for additional details.

\vspace{2cm}

{\bf References }\\

[Ch]\quad  
 Chouikha A R, \ On the expansions of elliptic functions and applications, in 
Algebraic methods and q-special functions, Editors:  
 C.R.M. Proceedings  and Lectures Notes, A.M.S., vol 22,   Providence, 1999, 53-57.\\
 
[Ch1]\quad 
 Chouikha A R, Note on trigonometric expansions of theta functions, {\it  J. of Comp. Appl. Math.} {\bf 153} (2003),  119-125.\\

[Ch2]\quad 
 Chouikha A R, Trigonometric expansions of theta functions and applications, 
Math. ArXiv, http://front.math.ucdavis.edu/math.NT/0112137.\\

[B]\quad  Erdélyi A, Magnus W, Oberhettinger F, Tricomi F,  Higher transcendental
functions  Vol. I and III. Based on notes left by H. Bateman, Editors: Robert E. Krieger Publish. Co.,
Inc., Melbourne, Fla., 1981.\\

[K-L-S]\quad 
 Khare A, Lakshminarayan A, Sukhatme U,  
Cyclic Identities Involving Jacobi Elliptic Functions. {\it J. Math. Phys.} {\bf 44, n 4} (2003), 1822-1841.\\

[L]\quad  Lawden D F, 
 Elliptic functions and applications, 
Springer-Verlag, New-York, 1989.\\

[M]\quad 
 Mumford D, Tata lectures on theta. vol II. Jacobian theta functions and differential equations, With  C. Musili, M. Nori, E. Previato, M. Stillman and H. Umemura. Progress in Mathematics, {\bf 43}, Birkhäuser, Boston, 1984.\\

[W-W]\quad 
 Whittaker E T and Watson G N, A course of Modern Analysis, Cambridge University Press, Cambridge UK, (1963).

\end{document}